\documentclass{sigchi}

\usepackage{balance}       
\usepackage{graphics}      
\usepackage[T1]{fontenc}   
\usepackage{txfonts}
\usepackage{mathptmx}
\usepackage{url}
\usepackage{breakurl}

\usepackage{color}
\usepackage{booktabs}
\usepackage{textcomp}

\usepackage{microtype}        
\usepackage{ccicons}          

\usepackage{todonotes}
\usepackage{xcolor}
\definecolor{mypink1}{rgb}{0.858, 0.188, 0.478}
\newcommand{\wenwen}[1]{\textcolor{black}{#1}}
\newcommand{\ali}[1]{\textcolor{black}{#1}}
\newcommand{\ryan}[1]{\textcolor{black}{#1}}

\def\plaintitle{Du Bois Wrapped Bar Chart: Visualizing categorical data with disproportionate values}
\def\plainauthor{Alireza Karduni$^{1}$, Ryan Wesslen$^{1}$, Isaac Cho$^{1,2}$, Wenwen Dou$^{1}$}

\def\plainkeywords{bar chart; graphical perception; user study; evaluation; Mechanical Turk; crowdsourcing; Information visualization }

\makeatletter
\def\url@leostyle{%
  \@ifundefined{selectfont}{
    \def\UrlFont{\sf}
  }{
    \def\UrlFont{\small\bf\ttfamily}
  }}
\makeatother
\def\pprw{8.5in}
\def\pprh{11in}

\definecolor{linkColor}{RGB}{6,125,233}

\begin{document}

\title{\plaintitle}
\author{\plainauthor\\
 \affaddr{$^{1}$University of North Carolina at Charlotte, $<$akaduni, rwesslen, wdou1$>$@uncc.edu}\\ 
\affaddr{$^{2}$North Carolina A\&T State University,  icho@ncat.edu}
}

\maketitle

\begin{abstract}

We propose a visualization technique, Du Bois wrapped bar chart, inspired by work of W.E.B Du Bois.  Du Bois wrapped bar charts enable better large-to-small bar comparison by wrapping large bars over a certain threshold. We first present two crowdsourcing experiments comparing wrapped and standard bar charts to evaluate (1) the benefit of wrapped bars in helping participants identify and compare values; (2) the characteristics of data most suitable for wrapped bars. In the first study (n=98) using real-world datasets, we find that wrapped bar charts lead to higher accuracy in identifying and estimating ratios between bars. In a follow-up study (n=190) with 13 simulated datasets, we find participants were consistently more accurate with wrapped bar charts when certain category values are disproportionate as measured by entropy and H-spread. Finally, in an in-lab study, we investigate participants' experience and strategies, leading to guidelines for when and how to use wrapped bar charts. 

\end{abstract}




\section{Introduction}

The concept of a bar chart was first introduced by French scientist Nicole Oresme in the 14th century \cite{beniger1978quantitative,der2014handbook}. However, many sources attribute the wide adoption of bar charts to the 1786 seminal work by William Playfair on ``Exports and Imports of Scotland to and from different parts for one Year from Christmas 1780 to Christmas 1781'' \cite{spence2006william,anyChart:online}. 

\begin{figure}[t]
  \centering
    \includegraphics[width=1.0\columnwidth]{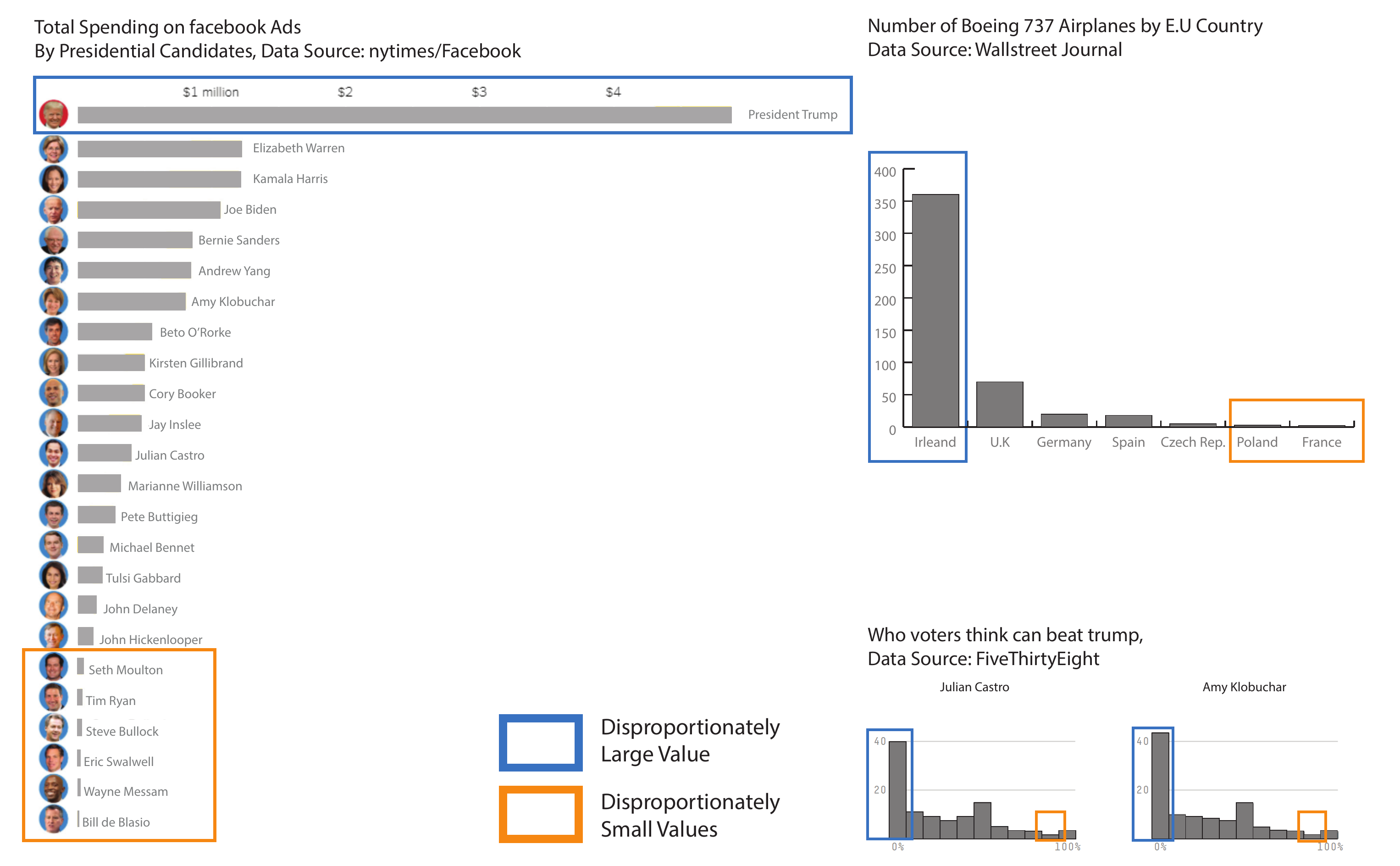}
  \caption{Three recreated examples of bar charts based on data and charts in news articles \protect\cite{NYtimesDatasets:online,WSJDatasets:online, 538Datasets:online}. All three charts are characterized by one extremely disproportionate value, leaving the smaller values difficult to estimate and compare.}
  \label{fig:examples}
\end{figure}

The modern definition of a bar chart is ``a chart that presents categorical data with rectangular bars with heights or lengths proportional to the values that they represent'' \cite{VisageBarChart:online}. Bar charts are now considered one of the most popular and prolific visualization techniques for communicating categorical values \cite{spence1991displaying,Zhao:2019:NPB:3290605.3300462}. However, certain data characteristics such as disproportionally large and small values make performing certain tasks with bar charts challenging. 
Figure \ref{fig:examples} shows three examples we identified from news articles \cite{NYtimesDatasets:online,WSJDatasets:online, 538Datasets:online} that use bar charts to visualize categorical values\footnote{\wenwen{We recreated the charts from the original data due to copyright issues.}}. All three examples are characterized by one disproportionately large value, which induces a significant white space-to-data ratio in the chart's plane. Estimating the smallest values and the ratio of largest to smallest values is challenging with charts like these. To address these challenges, we propose a new technique called a ``wrapped bar chart''.

\begin{figure}[t]
  \centering
    \includegraphics[width=1.0\columnwidth]{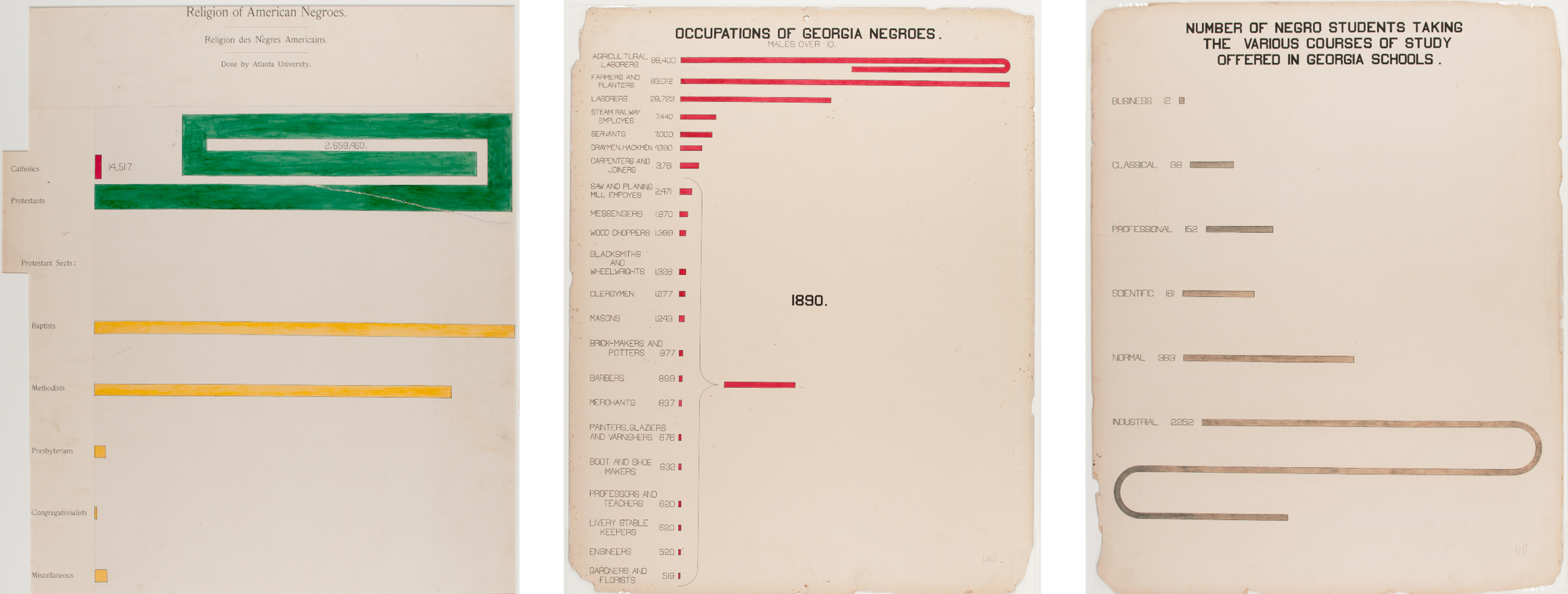}
  \caption{Three examples of wrapped bar charts designed by Du Bois \protect\cite{WEBDuBoi:online}. Images assessed from Library of Congress. Images courtesy of Library of Congress, Prints \& Photographs Division, LC-USZ62-1234 }
  \label{fig:dubois}
\end{figure}

Our work is inspired by the work of William Edward Burghardt ``W. E. B.'' Du Bois, a sociologist, historian, activist, and author who was also a prolific graphic designer \cite{WEBDuBoi:online}. His work for the 1900 Exposition Universelle in Paris utilized hand-drawn data visualizations to focus on the African American population in the 1890's South \cite{battle2018web}. Du Bois explored many different types of visualization techniques including using bar charts to highlight inequities for African Americans in Georgia. For example, African Americans worked disproportionately in agriculture occupations such as laborers, farmers, and planters but rarely in manufacturing and professional occupations like engineers, masons, merchants, and barbers. This extreme difference made drawing African-American occupation counts with a standard bar chart to be ineffective, especially when comparing the largest occupation (bar) to the smallest occupation (bar). To address this problem, he introduced an innovative solution to his bar charts: to wrap the tallest bars around themselves, allowing more room to show smaller bars. Figure \ref{fig:dubois} presents three charts by Du Bois that include a ``wrapped'' bar for its largest data value. 

Present-day visualization practitioners attempt to deal with the problem of disproportionate categorical data in bar charts through ad hoc solutions that have known issues. For example, one common solution is ``breaking the axis'' by interrupting the vertical axis of a chart at a specific point through a break. Similarly, cut-off bars limit the upper bound of the axis. However, these techniques are known to mislead users \cite{bryan1995seven}. 
While the two aforementioned techniques use a linear scale, other studies have analyzed scientific data with a non-linear scale (e.g. logarithmic) when data values cover a large range of magnitudes \cite{borgo2014order}. 
However, to our knowledge Du Bois' wrapping technique has not been studied by the visualization community to evaluate its effectiveness. Our work is to explore the potential and usages of Du Bois' intriguing innovation. To this aim, we first introduce a technique for developing wrapped bar charts in web-based visualizations. Second, we conduct three experiments to explore the limitations, potentials and benefits of wrapped bar charts in helping users conduct data analysis tasks in datasets with large variance between data values. Our paper makes the following contributions:


\begin{itemize}
    \item We design and develop the Du Bois wrapped bar chart\footnote{The similar term ``wrapped bar graph'' has been used for a different visualization technique. Stephen Few \cite{Few2013} introduced this term for a chart that splits the sorted bars in a horizontal bar chart into multiple columns to eliminate the need for scrolling \cite{Niklas2019}. In contrast, the Du Bois wrapped bar chart wraps disproportionately large bars so that small values are discernible. 
    } inspired by Du Bois' work for web-based visualizations using D3.js.
    \item We conduct two online crowd-sourced experiments for comparing participants' performance with wrapped bar charts versus standard bar charts in identification and ratio estimation tasks, using accuracy and time spent to complete each task as metrics of performance. 
    \item We conduct an in-lab focus group study that focuses on gathering information about participants' strategies when using wrapped bar chart, as well as collecting suggestions to improve the wrapped bar design.

\end{itemize}


In the following sections, we review related work, design of wrapped bar charts, experiment design, discussion of results, and conclude with limitations and future works.

\section{Related Work}\label{RelatedWork}

\subsection{Graphical Perception Studies of Bar Charts}
Many empirical studies have been conducted on understanding users' perception of the visual encoding of bar charts \cite{cleveland1984graphical, heer2010crowdsourcing}. Findings from these experiments provide design guidelines and considerations for applying bar charts to various tasks such as difference comparison and proportion estimation. The bar chart experiment by Cleveland \& McGill studied participants' estimate of the proportion of the lengths of two bars \cite{cleveland1984graphical}. The experiment included 10 different pairs of bars with proportion values ranging from 17.8\% to 82.5\%. Participants' responses were measured by the average log absolute error - the difference between a participant's answer and the true percentage value. The analysis results revealed that the accuracy was affected by multiple factors: the responses were more inaccurate with the stacked bar charts than standard bar charts; the responses were most inaccurate for true percents around 60\% - 80\%; and the accuracy decreased with increasing space between the two bars. 

As an effort to study crowdsourcing for perceptual experiments, Heer \& Bostock conducted an approximate replication of Cleveland \& McGill's bar chart study with slight modifications on the true percents \cite{heer2010crowdsourcing}. The new results are consistent with the findings from the original study although the log absolute error results are better in the new study, possibly due to the true percents being rounded to whole numbers instead of numbers with decimals. 
More recently, through a series of follow-up experiments, Talbot et al. \cite{talbot2014four} aimed to explore the results from Cleveland \& McGill's experiment in order to explain the sources of bar chart interpretation error. Their findings -- including evidence that shorter bars are more difficult to compare -- suggest that additional studies are needed to provide a more complete understanding of what impacts (bar) chart perception. \wenwen{Talbot et al. also found that the separation (gap) between bars increase bar comparison difficulty but the effect of intervening distractor bars are small.} More recently, Zhao et al. \cite{Zhao:2019:NPB:3290605.3300462} conducted experiments to evaluate how the perception of a bar changes based on the heights of its neighboring bars~ and found the neighborhood effect does exist.

Our study is inspired by the design and findings from the aforementioned graphical perception experiments. More specifically, the tasks in our study include the ratio estimate and identification tasks that require comparison of bars appearing both similar and orders-and-magnitude different in heights. \wenwen{A notable difference between the ratio estimation task in this study and previous studies on bar charts is that we chose to ask participants to calculate the ratio of largest/smallest bars as opposed to the percentage of smallest/largest bars. 
While the smallest percentage (small/large) evaluated by Cleveland et al. and Heer et al. was 17.8\%, our study would involve estimating percentages as small as 0.2\% and 0.24\% if framed as small/large estimation seen in the previous studies.} 

\subsection{Evaluating Alternative Bar Chart Designs}
Multiple studies have evaluated alternative bar chart designs. Skau et al. \cite{skau2015evaluation} introduced different illustrative embellishments which are design alternatives of a (single-series) bar chart. Skau et al.'s alternative bar chart designs drew inspiration from visual embellishments \cite{borgo2012empirical} which provide non-linguistic rhetorical figures that are used frequently in the visual/performing arts, advertisements, graphical user interfaces, etc. Srinivasan et al. \cite{srinivasan2018s} introduced design variants of the multi-series bar chart. Both studies evaluated the performance on the task of comparing values between bars. Both studies added additional information to single- and multi-series bars such as embellishments \cite{skau2015evaluation} and difference overlay \cite{srinivasan2018s}. However, such additions do not affect how the bars are plotted.

\textbf{Y-axis Distortion}. In comparison, another suite of alternative bar chart designs distort the Y axis in a bar chart thus affects how the length of the bars are determined and visually presented. 
Cut-off bars, scale break \cite{carvalho1992graphic}, and logarithmic scale are methods that were proposed to address problems classic linear bar charts may have in displaying disproportionately large values. These methods distort the Y-axis with a either non-linear scale or omitting a value range on the Y-axis. 
Cleveland discussed some of the techniques involving Y-axis distortion and called for experimentation for improving graphical communication in science \cite{cleveland1984graphicalOp}. Hlawatsch et al. noted that these distortions introduce a ``lie factor'' \cite {tufte1983vol} when there is discrepancy of effect size between the data and its representation \cite{hlawatsch2013scale}. The distortions will result in misleading scenarios for quantitative comparisons, which is the most common task for classic linear bar charts \cite{bryan1995seven}. More recently, Borgo et al. \cite{borgo2014order} proposed and evaluated an alternative bar chart design called Order of Magnitude Marker (OOMM) to facilitate the task of large magnitude number detection. The OOMM technique used a normalized scientific notation $A\times B^{10}$, with B determining the Y-axis scale and interval. The authors reported an empirical study (N=21) that demonstrates OOMM outperformed linear and log-scale bar charts design in identification and ration tasks. Note that the study did require participants to have basic knowledge of calculus and familiarity with concepts such as graphs and logarithmic scale. Therefore, we contend that OOMM and other techniques that distort the Y-axis may be better suited for communicating scientific data but may not be as effective for general audience \cite{cleveland1984graphicalOp,carvalho1992graphic}. In contrast, the wrapped bar chart was originally invented for exhibitions and aimed to communicate discrepancy between values to general audience. For these reasons, our experiments evaluated the proposed wrapped bar chart technique against  standard linear bar chart, without considering techniques that distort the Y-axis. Since the wrapped bar chart employ the same linear scale, we refer to standard linear bar chart as standard bar chart when reporting the experiments.



\section{Design of Du Bois Wrapped Bar Charts}\label{design}

To develop wrapped bar charts inspired by Du Bois, we define two threshold variables that determine where and how to wrap a bar. First, we define a threshold $t1$ that determines where on the axis to wrap a bar for the first time. For a value exceeding $t1$, the bar representing this value will be wrapped. In terms of the wrapping direction, we decided to wrap down-to-up 
(first example in Figure \ref{fig:dubois}). This design choice allows room for wrapping multiple times and ease of estimating the length of the wrapped portions. 

\begin{figure}[t]
  \centering
    \includegraphics[width=1.0\columnwidth]{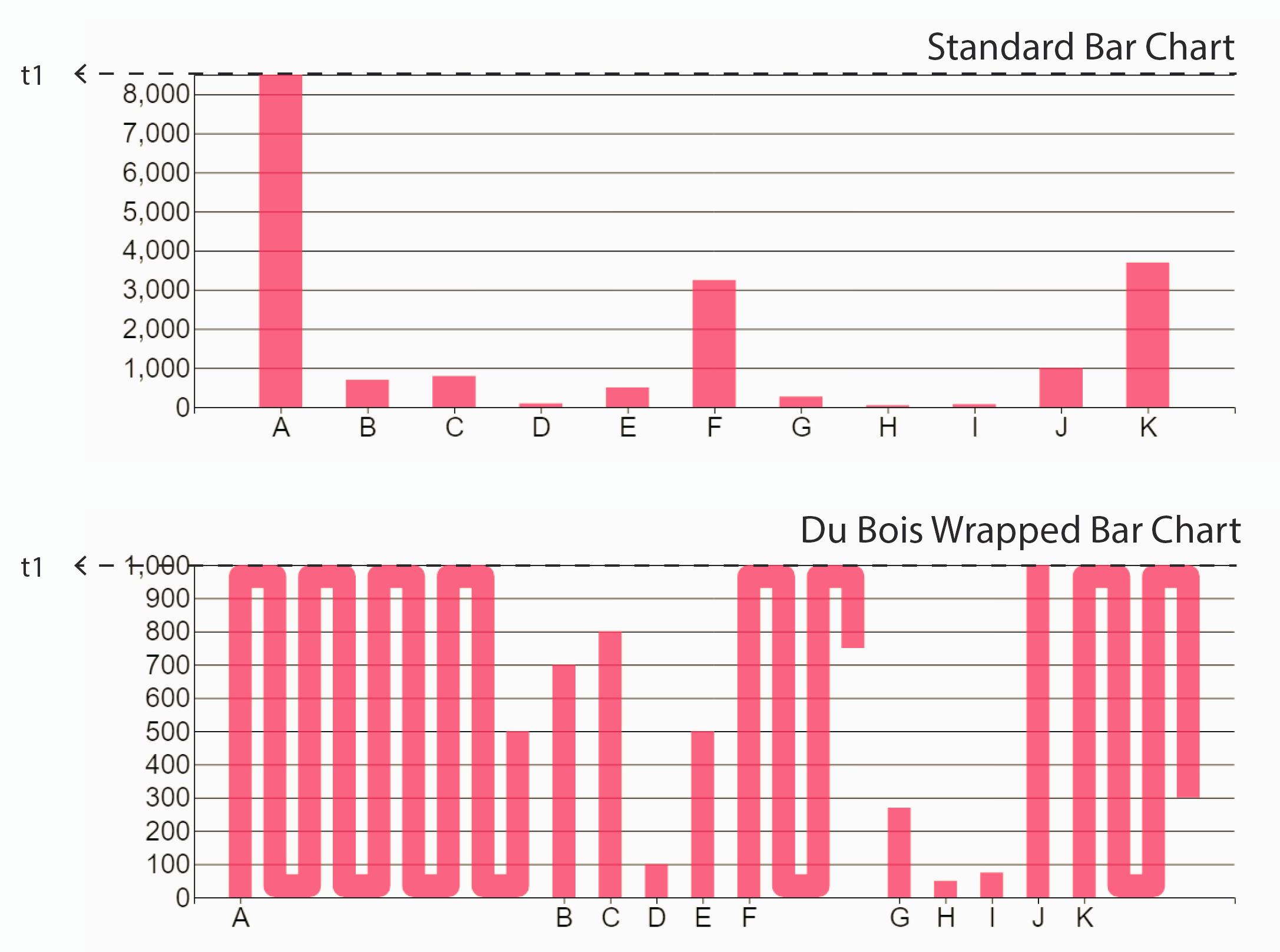}
  \caption{Top: Standard bar chart. Middle: Wrapped bar chart at full length. Bottom: Wrapped bar chart at half length. Color choices for the bars and background are adopted from Du Bois' work in Figure \ref{fig:dubois}.}
  \label{fig:implementation}
\end{figure}

\ali{
Essentially, $t1$ limits the numerical axis by a value and wraps any bar that exceeds that value. In Figure \ref{fig:implementation}, $t1$ is set to $1000$. As a results, the largest bar with a value of 8500 has 8 full wraps (8 times $t1 = 1000$ equals 8000) with the tail reaching half of the numerical axis (equaling 500). 
}

When designing the wrapped bars, we made multiple design decisions. First, in addition to a one-to-one ratio between the width of bars and the gaps between categories, we also determined on half of a bar-width gap for separating bars in the wrapped portion. Second, in our design, the value of a wrapped bar is the sum of only the vertical lines in a vertical wrapped bar charts and the wrapping portions in the horizontal axis do not contribute to the total value. Similarly, the vertical wrapping portions in a horizontal wrapped bar chart would not count toward the total length. Therefore, our wrapped bar chart maintains a linear relationship between values and vertical height of bars. Such design decision allows for easy estimate of the overall value of a wrapped bar. For example, a bar with a value of 5,500 in a wrapped bar chart with $t1 = 1000$ and $t2 = 1$, wraps a total of 5 times with 500 as the tail of the wrapped bar. These charts are developed using D3.js and an interactive version of the wrapped bar chart can be viewed at \url{https://wrapped-barchart.herokuapp.com/}. 

Note that the wrapped bar chart prototype was first developed for Study 1 (shown in Figure \ref{fig:study1datasets}). We improved the design of the wrapped bar chart based on feedback collected through Study 1. The changes were mostly cosmetic, including changing the bar and background color to resemble Du Bois' original work, and adjusting the bar width and gaps as detailed above. The final bar charts design used in Study 2 are illustrated in Figure \ref{fig:implementation} (top and middle charts).


\section{Study 1: Pilot}
\subsection{Motivation and Hypotheses}
\label{study1}

Revisiting Du Bois' wrapped bar chart on African American students' enrolled courses (Figure \ref{fig:dubois}-right), we can see that the number of students enrolled in industrial courses has orders of magnitude larger than ones enrolled in business courses. From this example, we hypothesize that wrapped bar charts may bring two advantages when presenting datasets with high variance 1) the ability to maintain a fixed scale in a bar chart that allows for better visibility for bars with smaller values and 2) enabling more precise comparison/estimation of differences between larger and smaller values. However, we recognize that wrapping would discount the preattentiveness of the length of the bars \cite{treisman1985search}, thus tasks of identifying the highest bar and estimating the value may take longer. 

Based on these factors, we develop three hypotheses about wrapped bar charts in comparison to standard bar charts. Study 1 is conducted on two common types of tasks for a bar chart, namely \textbf{identification} and \textbf{ratio estimation} \cite{borgo2014order, srinivasan2018s, cleveland1984graphical,heer2010crowdsourcing, talbot2014four}. 
For these two types of tasks, we hypothesize that: 

\begin{itemize}
    \item \textbf{H1:} Participants will achieve higher accuracy with wrapped bar charts in identifying smallest values.
    \item \textbf{H2:} Participants will be more accurate with wrapped bar charts in estimating ratios involving smallest values.
    \item \textbf{H3:} Participants will take longer to complete of the tasks with wrapped bar charts, e.g., identifying the bar with the largest value, estimating the ratio of largest to smallest.
\end{itemize}

In addition to these hypotheses, we also want to measure the time performance of each participant per dataset to ensure that any gains in better task accuracy are not at the expense of additional time to complete. Therefore, we also report the time to complete the full set of tasks per individual per dataset.

\subsection{Experiment Design and Procedure}

\begin{figure}[t]
  \centering
    \includegraphics[width=1.0\columnwidth]{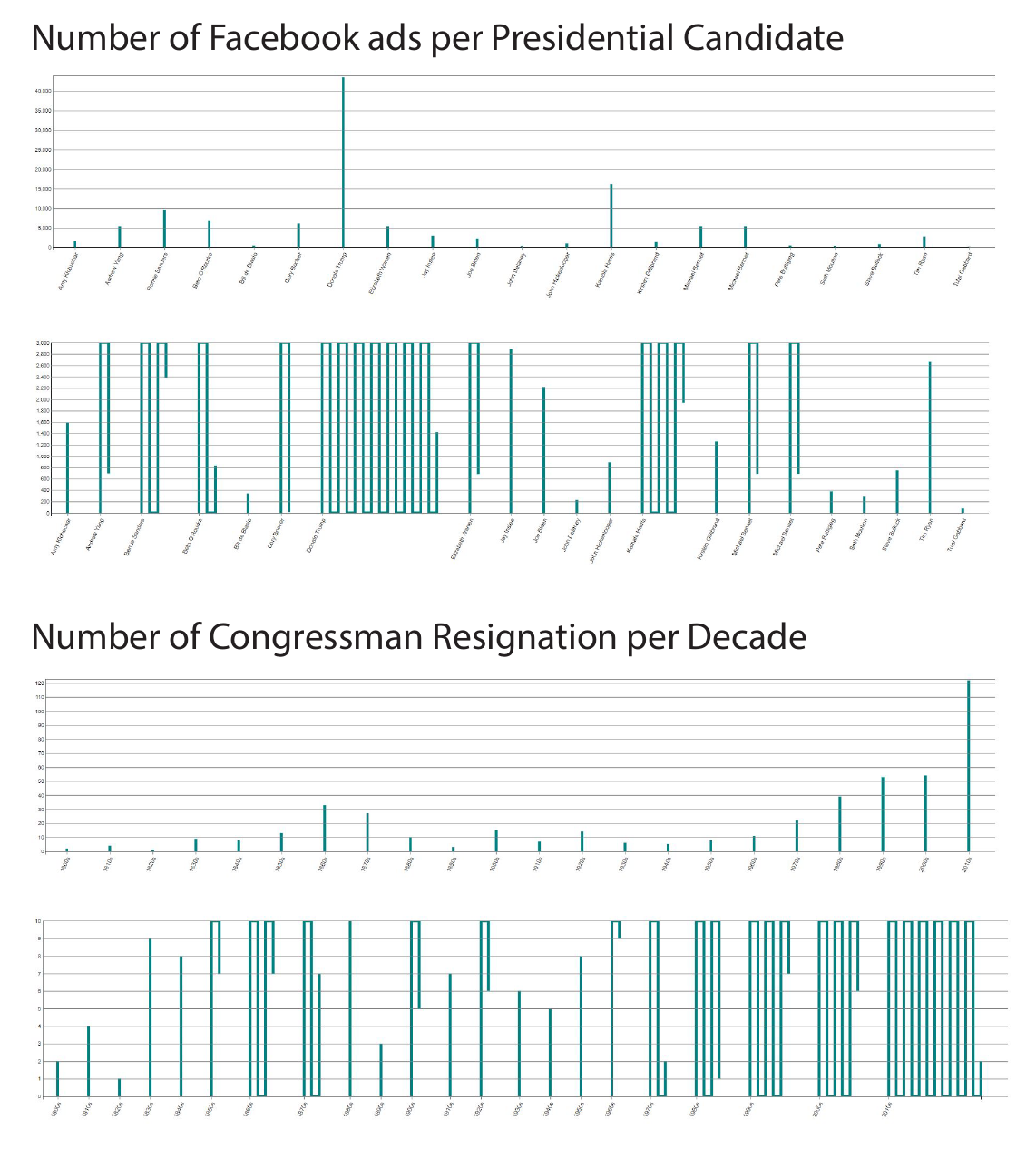}
  \caption{Two datasets used for Study 1. Top: Number of ad spending per presidential candidate, Bottom: Number of Congressional resignations per decade.}
  \label{fig:study1datasets}
\end{figure}

To compare participants' performance between wrapped bar charts and standard bar charts, we developed a web-based application for Study 1 
The application recorded participants' responses and response time. 
The experiment design involves two factors: 2 datasets $\times$ 2 chart types. We selected two real-world categorical datasets that exhibits very high variances between their largest and smallest categorical values (See Figure \ref{fig:study1datasets}). 
The differences in these data values make the task of identifying and comparing lowest bars very difficult with standard bar charts. The first dataset is about number of Facebook ads by US presidential candidates for the 2020 election (values ranging from 78 to 43,000), while the second dataset is about number of resignation of the members of the US Congress by decade (data ranging from 1 to 122).

The experiment has a between-subject design as each participant sees either a wrapped or standard bar chart for each dataset. As shown in Figure \ref{fig:studyProcedure}, each participant viewed both datasets but with varying bar chart types by being randomly assigned to either group A or B. Bars in Study 1 were plotted with a fixed bar width and gap width calculated to evenly distribute the bars in the charting area. (see Figure \ref{fig:study1datasets}). To catch participants that answered randomly, we added two simple bar charts (3-4 bars) before participants perform tasks on each dataset. 


Each participant is asked to complete six tasks for each bar chart presented to them during the study:


\begin{itemize}
    \item \textbf{T1:} \textbf{Identify} the bar with the \textbf{largest} value.
    \item \textbf{T2:} \textbf{Identify} the bar with the \textbf{smallest} value.
    \item \textbf{T3:} How many times is the bar with the \textbf{largest} value to the bar with the \textbf{smallest} value?
    \item \textbf{T4:} \textbf{Identify} the bar with the \textbf{second largest} value.
    \item \textbf{T5:} \textbf{Identify} the bar with the \textbf{second smallest} value.
    \item \textbf{T6:} How many times is the value of the \textbf{second smallest} bar to the \textbf{smallest} bar?
\end{itemize}

For the identification tasks (\textbf{T1}, \textbf{T2}, \textbf{T4} and \textbf{T5}), the participants used a mouse click to select a bar and then click on the "submit" button. For the ratio estimation tasks (\textbf{T3} and \textbf{T6}), the participants enter a number and submit their answer. 


\begin{figure}[t]
  \centering
    \includegraphics[width=1.0\columnwidth]{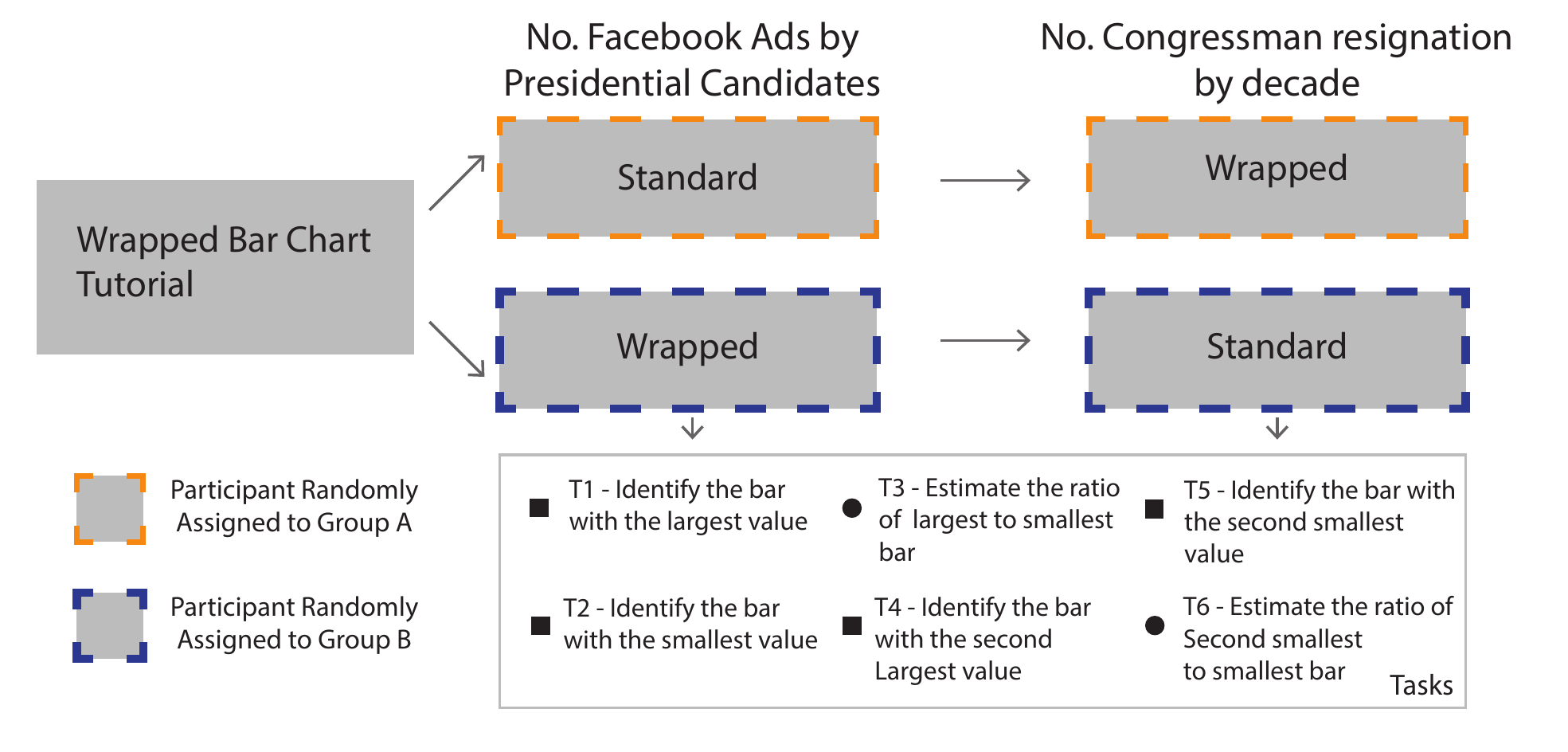}
  \caption{The general flow of Study 1. Each participant is assigned to one of the two (dataset*chart type) group after tutorial.}
  \label{fig:studyProcedure}
\end{figure}

\subsection{Experiment Results}

We deployed the study on Amazon Mechanical Turk. 98 participants completed the study with an average completion time of 5.5 minutes.\footnote{One participant was dropped for not completing and a second participant was dropped for repeating one of the sections.} Each participant was compensated \$0.50 for their time. In total, 56 participants were randomly assigned to Group A (standard bar first) and 42 to Group B (wrapped bar chart first).

We considered two different metrics to analyze the results. For \textbf{T1}, \textbf{T2}, \textbf{T4} and \textbf{T5} that resulted in categorical selection, we measured performance by user accuracy in identifying the correct bar value (e.g., largest, smallest). \ali{We define user identification accuracy as the percentage of times they identify a small or large bar correctly and report the estimate differences in percentage points}. For \textbf{T3} and \textbf{T6}, we used log absolute error similar to Cleveland and McGill \cite{cleveland1984graphical}. 

To analyze and report our experiment data, we use a non-null hypothesis statistical testing (non-NHST) approach focusing on sample means and bootstrapped 95\% confidence intervals (CIs) \cite{dragicevic2016fair}. \wenwen{To enable  comparison between our experiment results and other studies, we adopt the method by Talbot et al. to report estimates of simple effect sizes and associated CIs \cite{talbot2014four}, i.e. differences between estimated outcomes of experiment conditions followed by 95\% CIs in square brackets \cite{calmettes2012making}}. \ali{Narrow CIs compared to the estimated differences show strong evidence, while CIs that include zero imply more uncertainty about the sign of effect.} \wenwen{As highlighted by Kim et al., reporting mean differences as simple effect sizes has limitations as variations around these averages are not considered \cite{Kim2020}. To address such limitations, we also report Cohen's d, which is a measure of \emph{standardized} effect size \cite{cohen1988statistical}.} For consistency \wenwen{on the direction of reporting}, we will report \wenwen{effect sizes and CIs} of wrapped versus standard bar charts throughout this paper.

\begin{figure}[t]
  \centering
    \includegraphics[width=1.0\columnwidth]{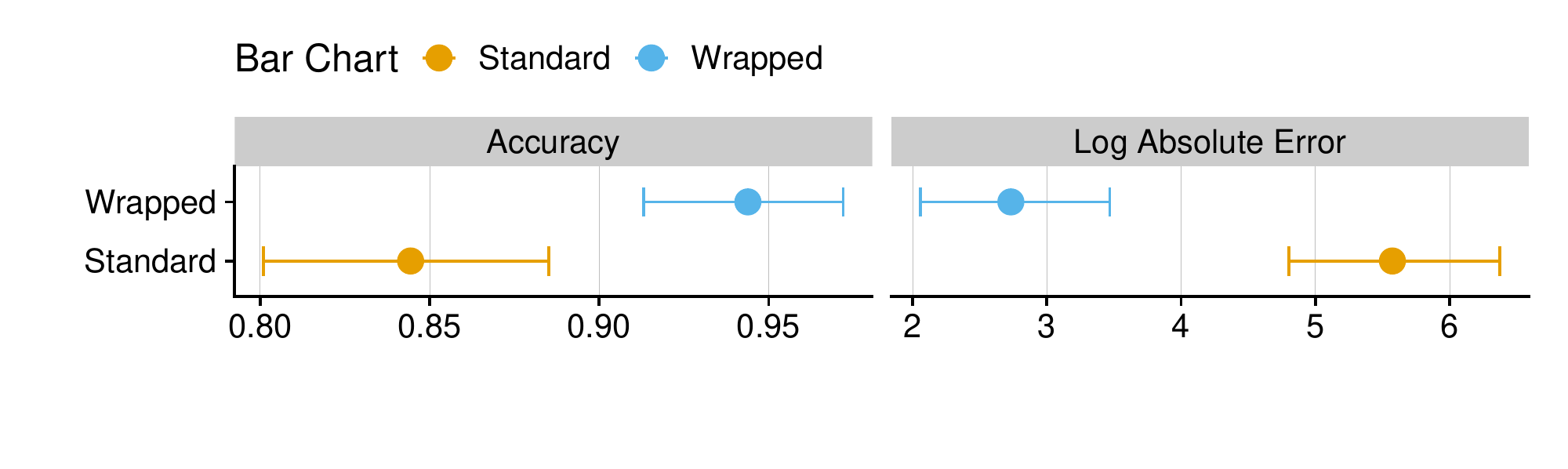}
  \caption{Study 1 mean and 95\% individual-level bootstrapped CI's (n = 98) for identification accuracy (left) and log absolute error (right) for participants using wrapped bar chart vs. standard bar chart.}
  \label{fig:overallaccuracy}
\end{figure}

As an overview, Figure \ref{fig:overallaccuracy}-left shows that participants achieved higher accuracy on the identification tasks by 10.07 [5.86,14.28] (d=0.32) percentage points.
Figure \ref{fig:overallaccuracy}-right shows that participants achieved higher accuracy using wrapped bar charts across two datasets on ratio estimation tasks as \ali{indicated by estimated log absolute error difference of -2.84 [-3.91, -1.78] (d=-0.52).}

\subsubsection{Identification Accuracy Wrapped vs. Standard Bar Chart}

\begin{figure}[t]
  \centering
    \includegraphics[width=1.0\columnwidth]{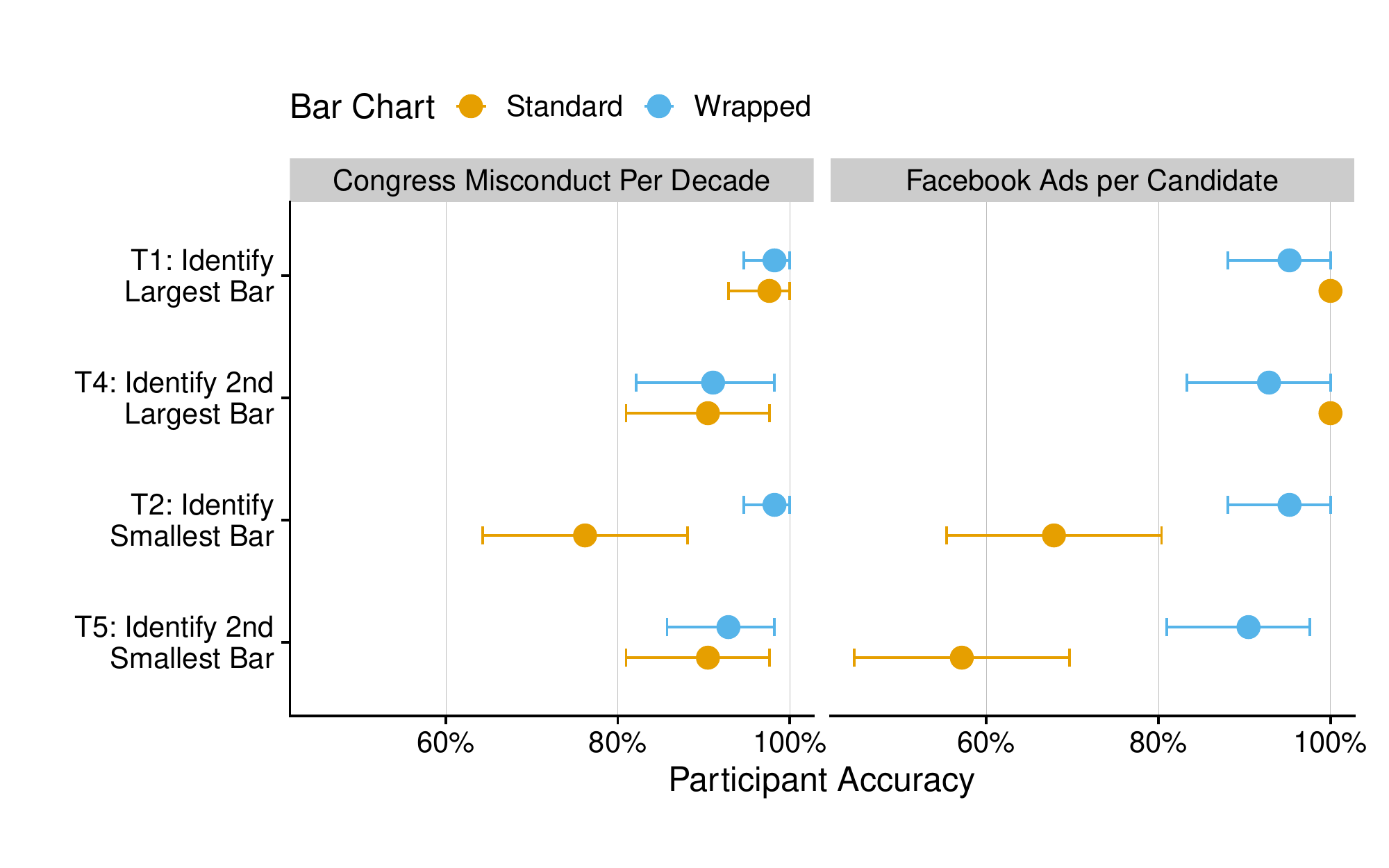}
  \caption{Study 1 mean accuracy and 95\% bootstrapped CI's (n = 98) for identification task accuracy (\textbf{T1}, \textbf{T4}, \textbf{T2}, \textbf{T5}) using a wrapped vs. standard bar chart.}
  \label{fig:taskaccuracy}
\end{figure}

Figure \ref{fig:taskaccuracy} shows the participants' accuracy on the identification tasks performed on both datasets. Consistent with \textbf{H1}, our results demonstrate the advantage of wrapped bar chart design in helping user identify bars with small values. Participants are more accurate with identifying the smallest value (\textbf{T2}) with wrapped bar chart design for both datasets. \ali{ For the Facebook Ads dataset, participants were more accurate on average by 27.38 [13.69, 41.07] (d=0.71) percentage points using wrapped bar charts. Similarly, for the Congress dataset, participants were generally more accurate by 22.61 [8.92, 36.30] (d=0.73) percentage points}. With the Facebook Ad dataset, participants are also more accurate at identifying the second largest value with wrapped bar chart design on average \ali{by 33.03 [17.26, 48.80] (d=0.78) percentage points. However, for the Congress dataset, we did not observe noticeable improvements for identifying second smallest 2.67 [-8.33, 13.69] (d=0.08).} \wenwen{The analysis did not reveal noticeable differences in identifying largest and second largest bars in both datasets across the two conditions.}


\begin{figure}[t]
  \centering
    \includegraphics[width=1.0\columnwidth]{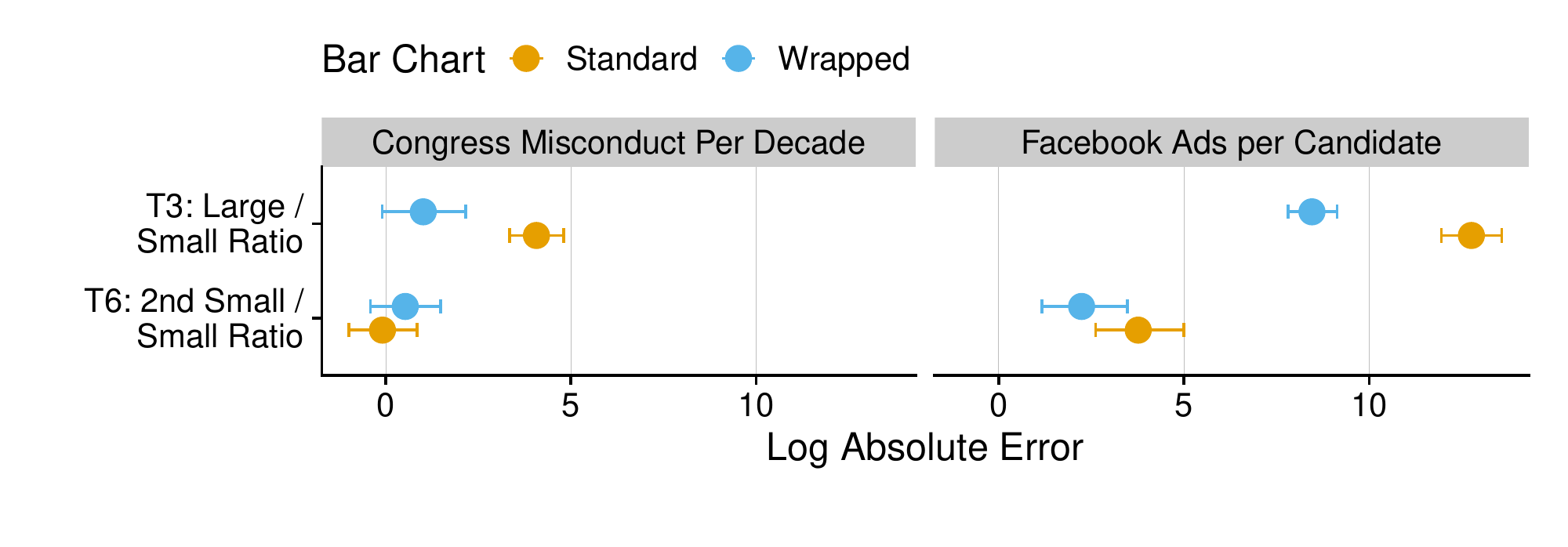}
  \caption{Study 1 mean and 95\% bootstrapped CI's (n = 98) for log absolute error (\textbf{T3}, \textbf{T6}) of participants using wrapped bar chart vs. standard bar chart across tasks and datasets.}
  \label{fig:logabsolutetask}
\end{figure}

\begin{figure}[t]
  \centering
    \includegraphics[width=1.0\columnwidth]{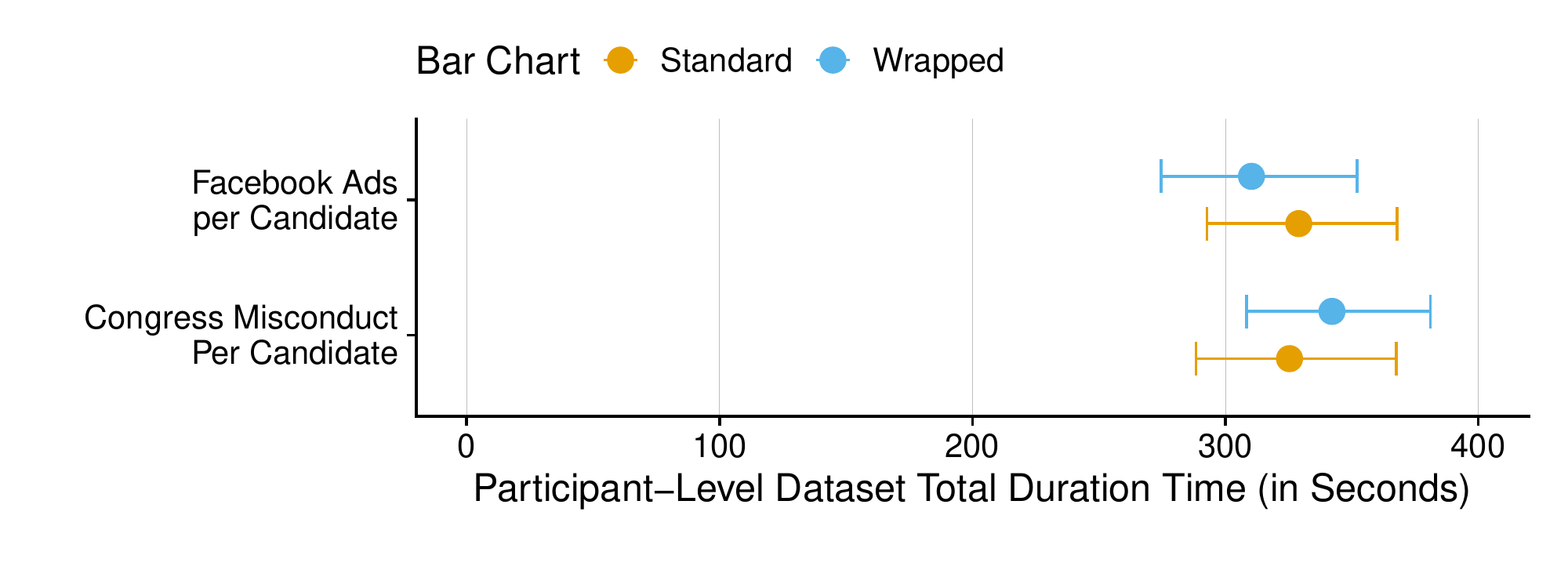}
  \caption{Study 1 mean completion time (in seconds) and 95\% bootstrapped CI's (n = 98) per participant for all six tasks comparing wrapped bar chart vs. standard bar chart.}
  \label{fig:times}
\end{figure}


\subsubsection{Ratio Estimation Accuracy Wrapped vs. Standard} 

Figure \ref{fig:logabsolutetask} shows the participant's performance on the ratio estimation tasks (\textbf{T3} and \textbf{T6}). The results for estimating the ratio of the largest to the smallest value (\textbf{T3}) is consistent with  \textbf{H2} that participants with the wrapped bar chart performed much better. \ali{ Specifically, the estimated mean difference in log absolute error is -4.30 [-5.36, -3.23] (d=-1.52) for the Facebook ads dataset and -3.01 [-4.38, -1.65] (d=-0.8) for the Congress dataset.} Inconsistent with \textbf{H2}, no significant difference is observed between accuracy using wrapped and standard bar chart to estimate the ratio of second lowest to the lowest value.


\subsubsection{Trial Completion Time} 

In Figure \ref{fig:times}, we provide the mean completion time and 95\% bootstrapped CI's per participant for each trial (i.e., dataset/bar chart combination with six tasks). We did not find evidence supporting \textbf{H3} as participants took nearly the same average time with wrapped bar chart as the standard bar chart for both datasets. 
However, one issue with only measuring the overall trial completion time is no visibility for the time taken for each individual task (e.g., largest bar identification, large / small ratio test). This issue is addressed in the second study.

\subsubsection{Discussion}
The results from the pilot demonstrated that wrapped bar chart design can lead to improved accuracy for identification and ratio estimation task. However, what kind of dataset could benefit from being visualized in a wrapped bar chart remains unclear. As a result, we conducted another study detailed in the next section. The design of Study 2 is also informed by lessons learned through the first study design, implementation, and analysis. For example, the issue raised when analyzing the trial completion time motivates an improved data collection for Study 2 to better track time duration by individual task.

\section{Study 2: Investigating when to use wrapped bar charts}
The primary goal of Study 2 is to investigate when it is advantageous to present data in a wrapped design instead of a standard bar chart. Study 2 is conducted to explicitly evaluate user performance on datasets of different characteristics.

\subsection{Improved Study Design and Hypotheses}


First, we highlight design changes from Study 2. Most importantly, to evaluate when to present data in a wrapped bar chart design, we leverage \textbf{theoretical metrics} (\textbf{normalized entropy} and \textbf{H-spread}) \wenwen{as heuristics} to determine what data characteristics made them ideal candidates for wrapped bar charts.

Given our Study 1 results, we improved our study design and implementation in the following ways:

\begin{itemize}
    \item Within-subjects study design to better measure the difference between standard versus wrapped bar charts.
    \item Remove identification and ratio test tasks of 2nd largest / smallest (\textbf{T4}, \textbf{T5}, and \textbf{T6}) to simplify study design and allow more repeated trials.
    \item Remove information about the dataset (such as number of Facebook ads in Study 1) to avoid distracting participants from focusing on the tasks. 
    \item Enhance time tracking to measure task-level duration rather than only trial-level duration.
\end{itemize}

\begin{figure}[t]
  \centering
    \includegraphics[width=1.0\columnwidth]{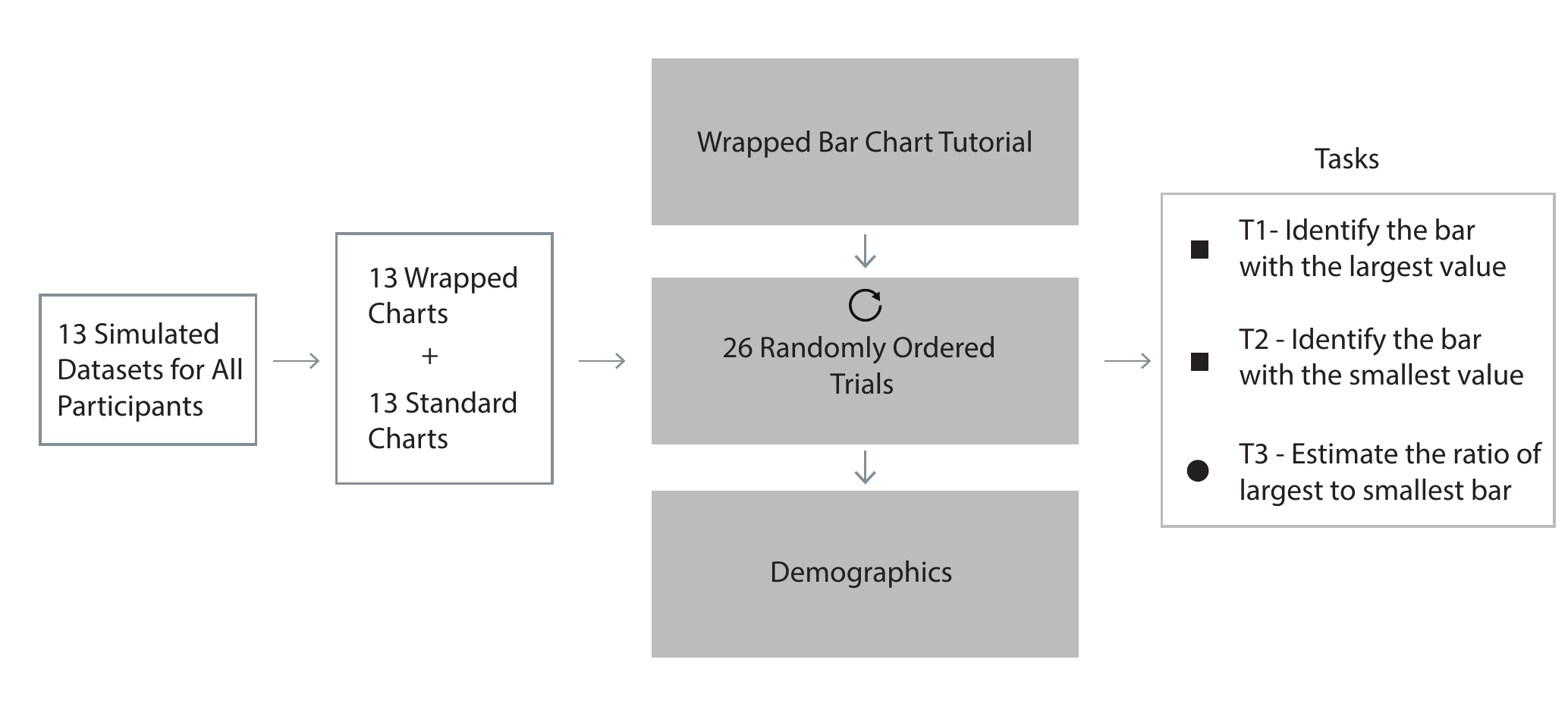}
  \caption{Study 2 Experiment Procedure}
  \label{fig:studyProcedure}
\end{figure}

Study 2 is a within-subjects, repeated trials experiment that measures the efficacy of a wrapped bar chart relative to a standard bar chart. The goal of Study 2 is to provide guidance for visualization designers when to consider using wrapped bar chart based on the characteristics of the data to be visualized. \wenwen{In particular, we propose two data metrics to quantify what data characteristics render wrapped bar charts useful. Entropy serves as the primary metric for characterizing data distribution. However, as shown in the bar chart simulations (Fig. \ref{fig:simulation}), each entropy bin can exhibit different top category concentration. We thus used H-spread as a secondary measure for how "far out" the highest bar is as defined by Tukey \cite{tukey1976exploratory}.}

\textbf{\wenwen{Primary} Data Metric: Information entropy to characterize the concentration of values.} In information theory, entropy measures 
the average (expected) amount of information from an event \cite{gray2011entropy}. For a more certain event, there is less information contained in that event and implies a lower entropy. Conversely, an event that is equally likely across all possible outcomes (e.g., a uniform distribution) will have high entropy. Essentially, entropy is inversely related to concentration. We hypothesize that participants will perform better with wrapped bar charts for our tasks on discrete categorical datasets with low entropy because such datasets exhibit a disproportionate concentration of values within a few categories. Conversely, such performance gains with wrapped bar charts will diminish for discrete categorical with higher entropy as their values are more evenly spread across categories. We define entropy as:

\[ \textrm{Entropy} = - \sum^{N}_{i = 1} {p_{i} * \log _{2} p_{i}} \]

where $p_{i}$ is the percent of values for a category $i$ and $N$ is the number of categories. One issue with entropy is that it increases with the number of categories, limiting the comparison across categorical datasets with a different number of categories. Therefore, we normalized entropy by dividing it by the log (base 2) of the number of categories ($N$). This normalization converts entropy into a range of values between 0 (all values concentrated into one bar) to 1 (uniform distribution).

\[ \textrm{Normalized Entropy} = \textrm{Entropy} / \log _{2} N \]

\textbf{\wenwen{Secondary} Data Metric: H-Spread to characterize disproportional values.} In his book, \textit{Exploratory Data Analysis}, John Tukey describes a method called `fences' for identifying values that are ``straying out far beyond the others'' in a dataset \cite{tukey1976exploratory}. His heuristic, which is widely used in boxplots to show disproportionate values, denotes the distance of a value away from the hinges (upper and lower quantiles) divided by the \textbf{H-spread} (the difference between the values of the hinges). He describes values larger than 1.5 times the H-Spread away from the hinges as `outside values'. Similarly, he names values that are 3 times the H-Spread outside the hinges as `far out' values.  Borrowing Tukey's heuristic, we formally define the ``H-Spread'' metric for a categorical dataset with $X$ values as: 

\[ \textrm{H-Spread} = (max(X) - Q3(X)) / (Q3(X) - Q1(X)) \]

Based on our results from Study 1 and our new theoretical metrics, we develop four hypotheses to test in Study 2.


\begin{itemize}
    \item \textbf{H4:} Participants will achieve a \textbf{higher identification accuracy} for the smallest values with wrapped bar charts for datasets with \textbf{low normalized entropy}.
    \item \textbf{H5:} Participants will achieve a \textbf{higher identification accuracy} for the smallest values with wrapped bar charts for datasets with \textbf{high H-spread}.
    \item \textbf{H6:} Participants will have \textbf{better large-to-small accuracy} (i.e., lower log absolute error) in estimating ratios between largest and smallest bars with the wrapped bar charts than standard bar charts.
    \item \textbf{H7:} Participants will spend \textbf{more time} with wrapped bar charts when estimating largest-smallest bars ratio.
\end{itemize}

\subsection{Data Simulation}

\begin{figure}[t]
  \centering
    \includegraphics[width=0.9\columnwidth]{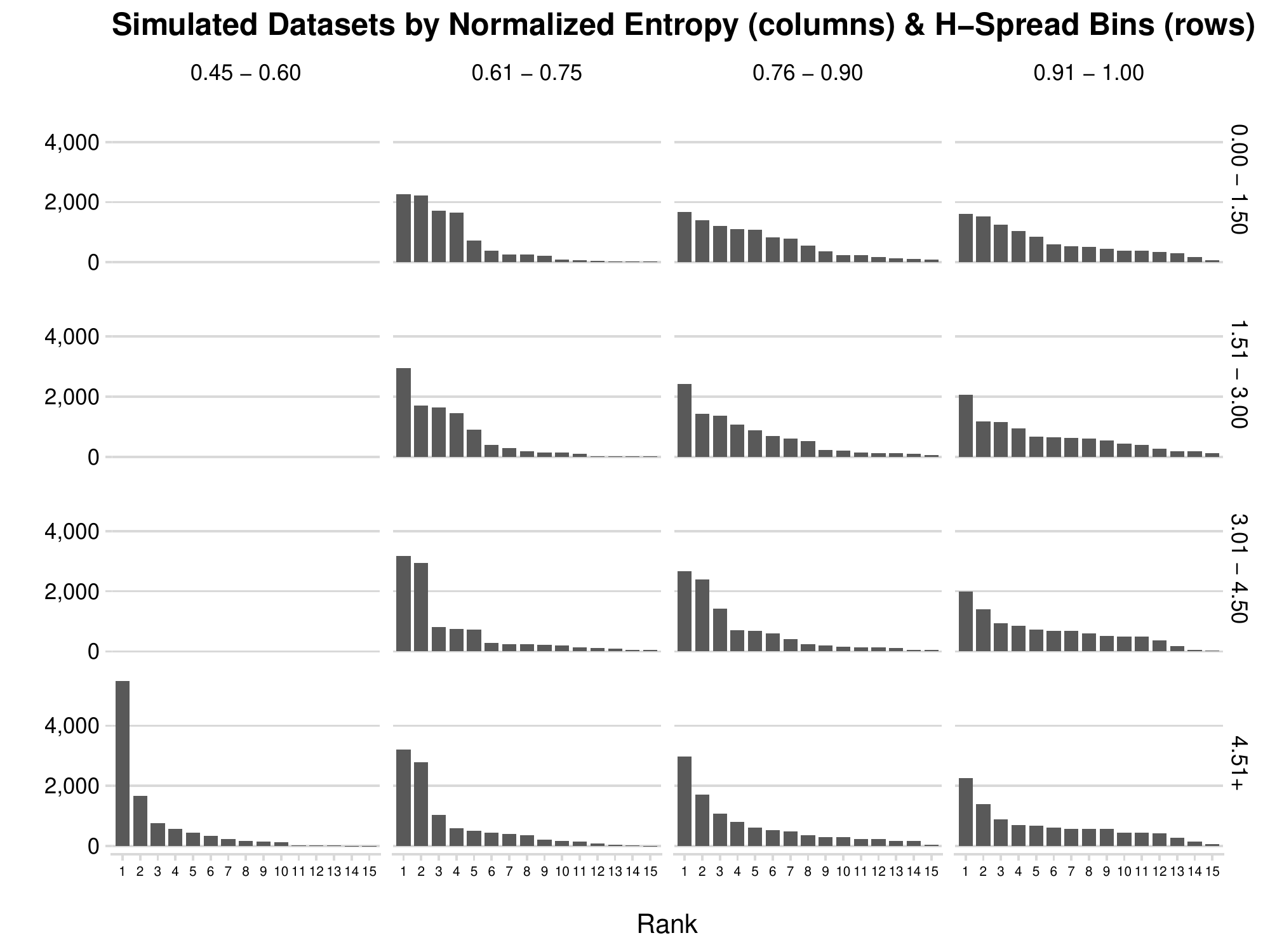}
  \caption{13 simulated datasets used in Study 2 that were randomly sampled from each Normalized Entropy (Columns) and H-Spread (Rows) bin combinations for a 10,000 dataset simulation.}
  \label{fig:simulation}
\end{figure}

To select our datasets for Study 2, we simulated 10,000 datasets based on 10,000 random draws from a fixed number of categories (e.g., 15 categories).\footnote{ Our simulator tool is a deployed R shiny app at \url{https://ryanwesslen.shinyapps.io/wrapped_bar_sim/} .} Next, we categorized each distribution into four fixed ranges (bins) for both normalized entropy and H-spread, creating a 4 $\times$ 4 grid of 16 possible combinations. We then randomly drew one dataset per normalized entropy and H-Spread bin combination to use in our experiment. Given we used only a finite number of simulations, we were only able to generate datasets for 13 of the 16 because datasets with a low normalized entropy (less than 0.60) and high H-Spread (more than 4.50) did not occur in our 10,000 dataset simulation. Figure \ref{fig:simulation} provides the 13 sampled datasets used in Study 2 by each of the bin combinations. Although each dataset is visualized as a sorted bar chart in the data simulation app, the bars appear in random order in the simulated dataset for Study 2.

\subsection{Experiment Design and Participants}

For Study 2, we designed a within-subject Amazon Mechanical Turk experiment in which participants were provided both the standard and wrapped bar charts in random order for each simulated datasets. We improved the design of standard and wrapped bars to maintain a 1 to 1 ratio between bar width and gaps. Given 13 datasets, each participant completed 26 total trials consisting of three tasks: (\textbf{T1}) identify the largest bar, (\textbf{T2}) identify the smallest bar, and (\textbf{T3}) ratio estimation of the largest-to-the-smallest bar value. We randomized the order of the dataset-chart combinations (trials) to reduce related confounding factors.

203 participants completed our study, each receiving \$2.00 reward for completion. On average, participants took 29 minutes and 42 seconds. We evaluated individual level performance on identification tasks to identify participants who performed unreasonably. After investigating the number of correct answers on the easiest task (i.e., \textbf{T1}), we dropped 13 participants who incorrectly identified the largest value for both standard and wrapped bar charts for more than 10 of 26 datasets, leaving 190 participants for Study 2.

\subsection{Experiment Results}

For reporting our results, we used individual-level bootstrapped means since participants completed repeated trials across each experiment factor. To calculate, we first averaged performance on a participant basis per factor (e.g., Entropy bin) and then bootstrapped on user level performance. This enabled us to control for heterogeneity between participant performance, which is commonly found in crowdsourced visualization experiments \cite{abdul2014repeated}. \ali{Since Study 2 is a within-subjects study, we report paired Cohen's d for within-subject paired samples \cite{lakens2013calculating}.}


\begin{figure}[t]
  \centering
    \includegraphics[width=1.0\columnwidth]{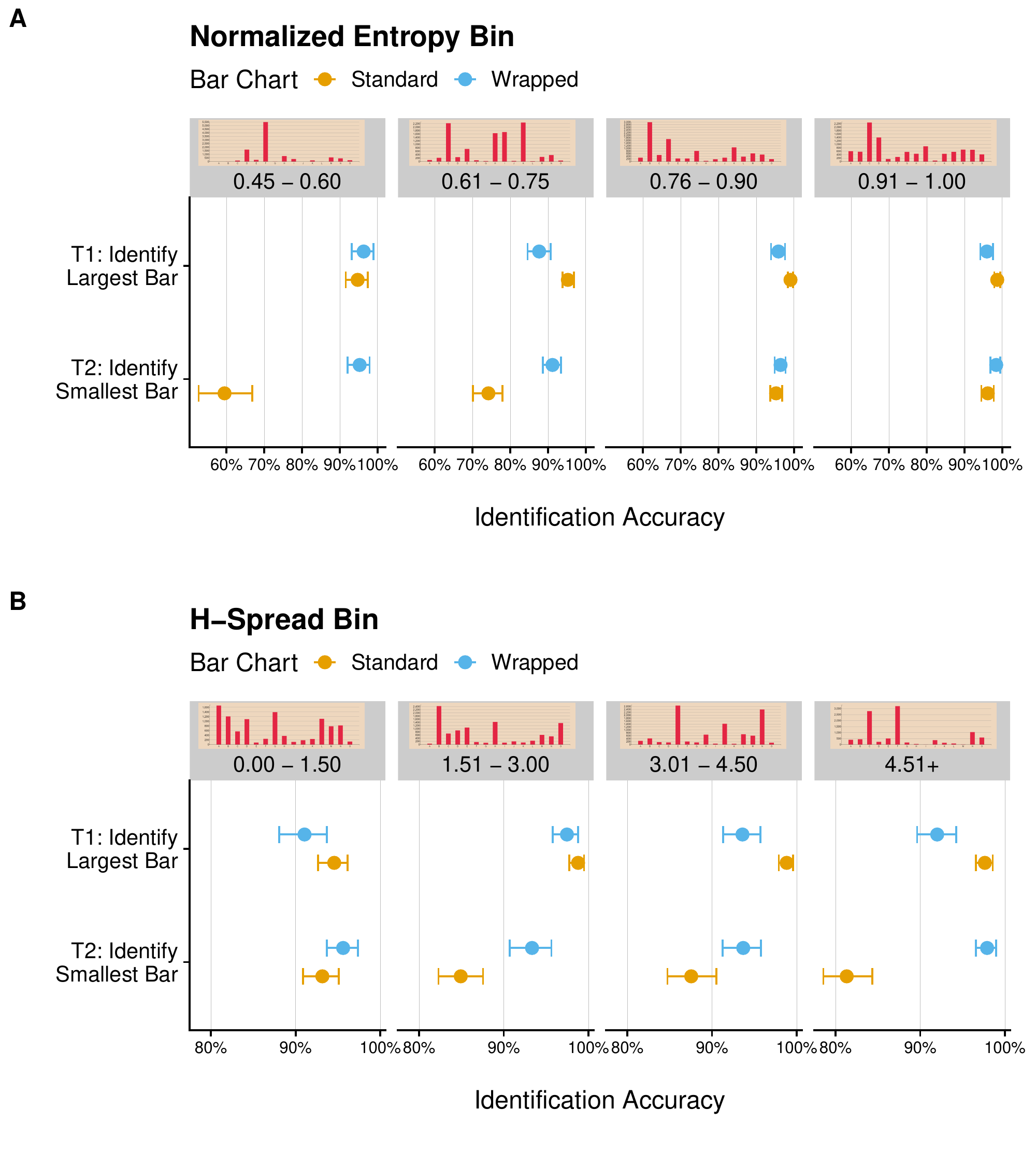}
  \caption{Study 2 large-small bar identification accuracy (\textbf{T1} and \textbf{T2}) and 95\% bootstrapped confidence intervals (n = 190) by Normalized Entropy bin (A) and H-Spread bin (B). We provide an example of a dataset from each bin as a visual cue.}
  \label{fig:accuracy}
\end{figure}

First, consistent with \textbf{H4} and \textbf{H5}, we find participants in general had similar or better small bar identification accuracy with a wrapped bar chart compared to a standard bar chart. Figure \ref{fig:accuracy} provides identification accuracy for Study 2 identification tasks (\textbf{T1} and \textbf{T2}) by normalized entropy and H-spread bin. The largest difference was in identifying the smallest value bars in datasets with normalized entropy below 0.75. \ali{We observed a mean difference of 35.78 [28.42, 43.15] (d=0.70) percentage points for wrapped bar charts in datasets with normalized entropy between 0.45-0.6 normalized entropy bin; and a difference of 16.88 [12.31, 21.44] (d=0.68) percentage points for 0.6-0.75 normalized entropy. For the entropy bins of 0.75-0.9 and 0.9-1, the identification accuracy mean differences are small, respectively 1.06 [-1.01 , 3.2] (d=0.08) and 2.01 [0.05 , 4.26] (d=0.2).}

In addition, we find that participants' had higher accuracy on small bar identification with wrapped bar charts than standard bar charts for datasets with H-spread larger than \ali{1.5. We observe the largest effect with the highest H-spread category of 4.5+, with accuracy difference of 16.53 [13.36. 19.71] (d=0.80) percentage points. For H-spread of 3-4.5 we observe a difference of 5.92 [2.06 , 9.78] (d=0.28) and for 1.5-3.0, a difference of 8.28 [4.64, 11.92] (d=0.45). We did not observe noticeable differences in datasets in H-Spread bin of 0-1.5.}

Interestingly, we find wrapped bar charts yield worse accuracy in 
largest bar identification for datasets with high H-Spread. \ali{For datasets with H-Spreads of 3.00-4.5, participants were on average less accurate by -5.28 [-7.67, -2.89] (d=-0.31) and for datasets with H-Spreads of 4.5+, a similar worse accuracy of -5.65 [-8.15, -3.15] (d=-0.34) percentage points when using wrapped bar charts.}
For normalized entropy, we observe small or no effect within different value ranges. For normalized entropy values between 0.45 and 0.6, we did not observe significant differences (d=0.05) in large bar identification accuracy between wrapped and standard bar chart. 
 \ali{For normalized entropy between 0.6-0.75, we observe on average a worse accuracy of -7.72 [-11.02, -4.42] (d=-0.36) percentage points with wrapped bar charts; for normalized entropy 0.75-0.9, we observe a mean difference of -3.26 [5.21, 1.31] (d=-0.22), and for 0.9-1.0 we observe a mean difference of -2.85 [-4.81, -0.89] (d=-0.22). }
We suspect that this may occur when multiple bars are wrapped, especially multiple times, leading to higher cognitive load in counting the number of wraps. We explore qualitative feedback to understand this hypothesis in our in-lab Study 3.


\begin{figure}[t]
  \centering
    \includegraphics[width=1.0\columnwidth]{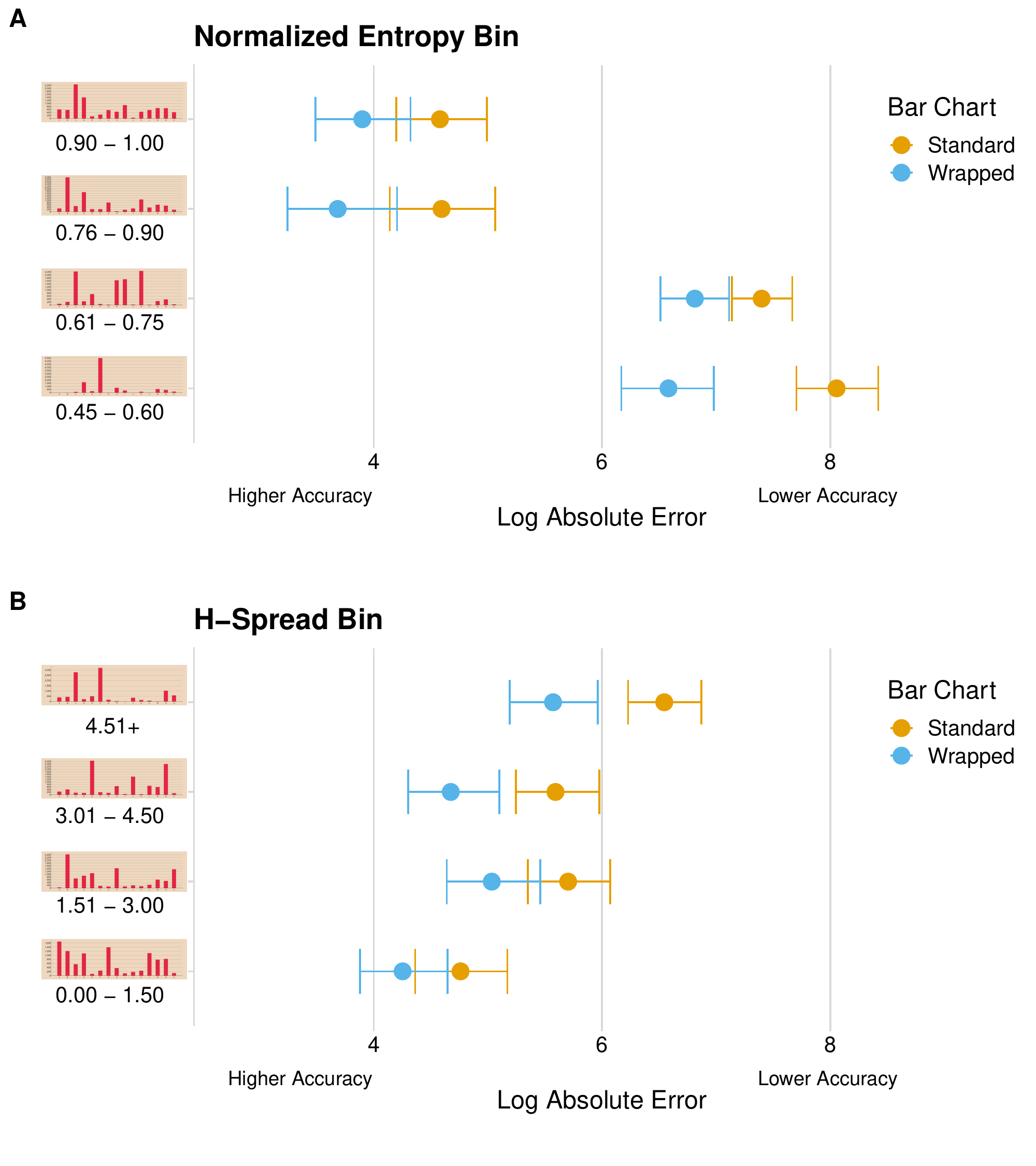}
    
  \caption{Study 2 the mean log absolute error and 95\% bootstrapped confidence intervals (n = 190) for large-small ratio estimation (T3) by Normalized Entropy bin (A) and H-Spread bin (B). We provide an example of a dataset from each bin as a visual cue.}
  \label{fig:error}
\end{figure}

Second, we observe evidence for \textbf{H6} that participants consistently have better large/small ratio accuracy with wrapped bar charts compared to standard bar charts. Figure \ref{fig:error} provides the mean log absolute errors and 95\% bootstrapped CI's by either normalized entropy or H-spread bins. \ali{For normalized entropy bins, we find the largest effect on datasets within the 0.45-0.6 range with a log absolute error mean difference of -1.47 [-2.01, -0.9] (d=-0.62). For normalized entropy within 0.6-0.75, we observe a mean difference of -0.58 [-0.97, -0.18] (d=-0.54). For normalized entropy of 0.75-0.9, we observe a mean difference of -0.89 [-1.54, -0.25] (d=-0.57). And we observe a mean difference of 0.67 [-1.26, -0.087] (d=-0.56) for normalized entropy values between 0.9-1.0.}

\ali{For H-Spread bins, we find the highest mean difference of -0.97 [-1.49, -0.46] (d=-0.69) in the H-Spread 4.5+ datasets. We observe a similar mean difference of -0.91 [-1.47, -0.34] (d=-0.55) for datasets with H-Spread in the 3-4.5 range. We observe a smaller effect of -0.67 [-1.21, -0.14] (d=-0.51) for H-Spread 1.5-3 datasets.}
In summary, we observe that participants achieved higher accuracy in  ration estimation in low normalized entropy  and high H-spread datasets. 


\begin{figure}[t]
  \centering
    \includegraphics[width=1.0\columnwidth]{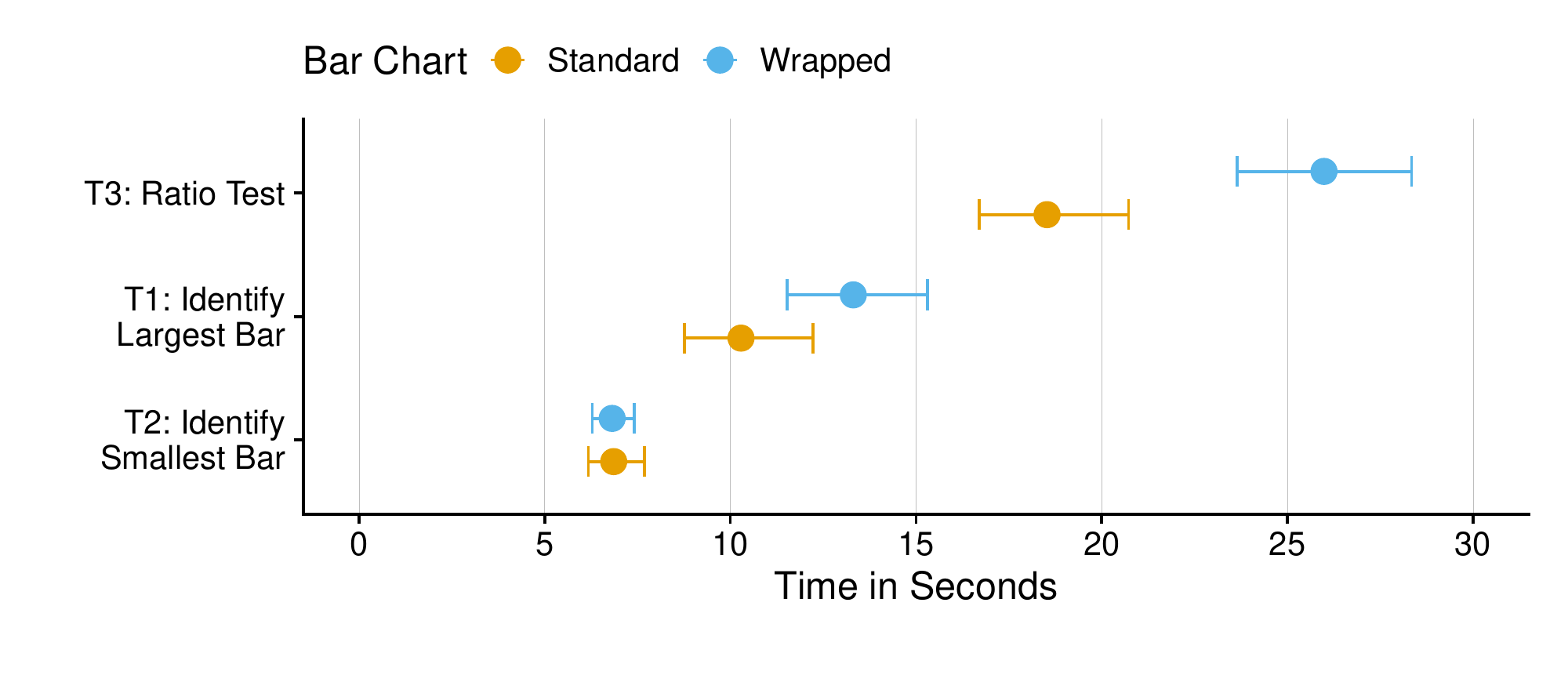}
  \caption{Study 2 mean completion time (n = 190) and 95\% bootstrapped confidence intervals for each task.}
  \label{fig:study2-time}
\end{figure}

Third, we find some evidence in support of \textbf{H7}
\ali{that on average participants tend to take 7.31 [4.98, 9.64] (d=0.42) seconds longer} 
to complete the ratio task (\textbf{T3}) with a wrapped bar chart as compared to a standard bar chart. We find a smaller mean difference in time for the wrapped bar chart in identifying the largest bar \textbf{T1} 
\ali{of 3.04 [0.64, 5.44] (d=0.18) additional seconds}
However, we find no difference in completion time for identification of the smallest bar \textbf{T2}. Figure \ref{fig:study2-time} provides the mean time to complete each task along with 95\% CI's. While we anticipated some increase in task time with the wrapped bar charts, overall, we do not find burdensome time effects that offset the benefits in identification and ratio accuracy.




\section{Study 3: Qualitative Evaluation} 

\subsection{Motivation and study design}
After obtaining the quantitative results in Study 1\&2, We design a qualitative study to gain a better understanding of users' individual experience with wrapped bar charts. We conducted in-lab focus group studies with a total of 24 participants. Our participants were graduate students enrolled in data science or architecture degree programs. None of these students had prior experience with wrapped bar charts. 
The procedure for Study 3 followed these steps: 
\begin{enumerate}
    \item Participants were given a brief overview of the study purpose (e.g., informed consent and instructions).
    \item Participants completed an online study similar to Study 2. However, to control for time, participants were asked to complete tasks on 14 of the 26 datasets.
    \item Participants completed an open-ended post-questionnaire individually to provide feedback on the study.
    \item Participants take part in a focus group session on user experience, challenges, and suggested improvements.
\end{enumerate}

The open-ended questions in the post questionnaire asked participants for their strategies on calculating ratios and their feedback on the usefulness of wrapped bar charts. After all participants in a session completed the study, we held group discussion about wrapped bars to further understand participants' feedback. We collected participants' written comments, transcribed audio recordings from the open discussion sessions. \ali{Two coders independently thematically analyzed users comments and feedback.} We identified four themes from participants' written comments and discussions:

\textbf{Benefits of wrapped bar charts:} The most prevalent themes from both focus group studies include the intuitiveness of the wrapped bar chart design, when they could be beneficial, and participants' strategies to interpret wrapped bars. Out of 24 participants, 21 mentioned they believe wrapped bar charts are useful for specific cases when some values in a dataset are disproportionate. One participant's comment summarizes this theme: \textit{``[Wrapped bar charts are useful] when there is an obvious disparate effect from one attribute over the others. If the axis scales were calibrated effectively, wrapped bar charts could more concisely convey differences in orders of magnitude''}. Another participant understood the benefit of wrapped bars but highlighted the importance of the tasks that chart designer want to explicitly support: \textit{``[Wrapped bar charts are useful] when it is hard to read minimums and when the minimums are crucial.''}.

\textbf{Usability of wrapped bar charts:} Participants highlighted two important points on the usability of wrapped bar charts. First, they understood that wrapped bars were useful for specific cases when largest values render a number of small values illegible. However, 16 participants argued that at some point the task of estimating values of the wrapped bars can be cumbersome, especially when there are many wraps. One participant commented: \textit{``wrapped chart is easier if there are only couple of wraps. but if number of wraps increase, it gets harder''}. A major source of annoyance for five participants was when they had to estimate the tail end value of wrapped bars stems from the opposite direction of the y axis (for the wrapper portion going top to bottom), as one participant mentioned: \textit{``... You need to do additional subtractions when the bar is coming downwards''}.

\textbf{Strategies for reading wrapped bar charts:} Another emerging theme in our discussions with participants was the different strategies they took to read wrapped bars, and calculate ratios between largest and smallest values. Majority of users simply tried to estimate the value of largest and smallest, and do a division task as described by two participants: \textit{``I assume a number for minimum between the range. Then, another number for the max. After that, I just divide and make the number round if possible.''} and \textit{``count the tall bar then the short bar and divide. for the tall bars I counted the number of full bars and multiplied by the max number then added the end bar up.''} However, three participants came up with another novel strategy to read wrapped bars: \textit{``I started estimating values of each of the largest and shortest columns [bars] and then divided them. But after a few iterations I found a better method. I estimated if stacking the shortest [bar] on itself X number of times would fill up to the first grid line and then counted how many lines the longest [bar] took up.''}  

\textbf{Suggestions for improving wrapped bar charts:} In the focus-group sessions, participants also discussed how to improve the design of wrapped bar charts. One suggestion was including an `inverse axis' to the right side of a wrapped bar chart to help estimate the value of the tail end of wrapped bars: \textit{``Use two y scales on both left and right side of the chart but in a reverse way''}. Moreover, participants had multiple suggestions on ways of simplify the estimation of value of largest bars by including count of wraps, using colors, and increasing gaps between wrapped bars.  

\section{Discussion and Future Work}

 In our two online experiments, we find specific cases (low entropy and high H-Spread) where users gain accuracy in identification and ratio tasks while using wrapped bar charts. However, we also identify subtleties and drawbacks in using wrapped bars in these specific cases. First, both quantitative and qualitative findings including participants' higher time spent on identifying the largest bar and mentions of annoyance with too many wraps provide evidence that wrapped bar charts are less preattentive than standard linear bar charts. Our participants recognized that for estimating values of wrapped bars, they need to go through a mathematical process of addition or multiplication of values. Based on the feedback in the focus-group study, we believe that this effect might be negligible for a small number of wraps. Adding information such as providing a label for the number of wraps is needed for larger number of wraps. 
 To understand the trade-off between cognitive load and accuracy in wrapped bar charts, we plan to conduct a future experiment studying the optimum threshold for wrapping bar charts and the limit to the number of times a bar can be wrapped before it becomes ineffective. 
 
 
 \wenwen{
 To understand the source of error w.r.t the ration estimation task, we further explored whether identification accuracy impacts users' ratio estimation accuracy. Out of 4,949 valid trial responses collected in Study 2, 12.5\% (617) inaccurately identified either the largest or smallest bar. \ryan{We ran a mixed effects model on only the trials with correct identification responses and found similar effects on log absolute error, with both means decreasing by about 0.1.} This exploration suggests an improvement for future study design, i.e. to explicitly highlight the correct highest and smallest bars for the ratio estimation task.}

In the future, there are several design improvements we plan to make and evaluate with wrapped bar charts in order to reduce the cognitive load required to interpret wrapped bars. First, as mentioned in the focus-group studies, we plan to introduce an inverse y-axis to the right side of a horizontal bar chart to help read the tail end of wrapped bars (when going from top to bottom). Moreover, we also plan to experiment different gap sizes between wrap bar charts to help with the task of counting the number of wraps. Third, we plan to study the effects of changing the second threshold of wrapped bars (see Figure \ref{fig:implementation} bottom) on participant performance. Finally, one solution to the trade-off between cognitive load and benefits of wrapped bars could be to introduce a new interaction technique for interactive bar charts. `Wrapping' would allow users to quickly move from standard bar charts to wrapped bar charts and observe the values of small bars. These design solutions require a comprehensive and systematic study to estimate and understand the benefits of these changes. 

From an application perspective, there are multiple extensions we can test with wrapped bar charts. First, in order to better understand the potentials of wrapped bar charts, we need to collect more data from users including mouse positions, user confidence level, and user satisfaction. Second, we need to compare wrapped bar charts with other types that are built to deal with datasets with values of high variance, such as broken axis and logarithmic scale. Third, during our development of wrapped bar charts, we developed a new kind of visual interaction by allowing users to interactively change the thresholds of wrapped bar charts ($t1\&t2$) and explore high and low data points. In a future study, we plan to thoroughly study this interaction technique for conducting similar visual tasks and learn about user preference. Last, we suspect that zoomed-out wrapped bar charts could work well with many more categories ($N$ > 50), like a power law distribution (e.g., Zipf's law for word counts) that is carefully ordered (e.g., by word-topic theme).

Finally, as a general suggestion for visualization designers to employ wrapped bar chart design within their work we provide the following recommendation resulting from our quantitative and qualitative evaluations: 
Using entropy and H-Spread as 
heuristics to measure a dataset and determine whether it is beneficial to use wrapped bar charts. We recommend using wrapped bar charts to visualize datasets with normalized entropy less than 0.75 as we see performance gain in identification and ratio tasks (Figures \ref{fig:accuracy} and \ref{fig:error}). Moreover, in line with Tukey's suggestion, when the H-Spread threshold is larger than 4.5 that includes far out values), wrapped bar charts could be a good design option.

\section{Conclusion}

In this paper, we designed, developed, and evaluated wrapped bar charts originally introduced by W.E.B. Du Bois to highlight the benefits of his visual innovation added to the standard linear bar charts presentation. We developed an implementation of wrapped bar charts for the web. We hypothesized that wrapped bar charts can outperform standard bar charts for datasets with disproportionate values. Using a web-based interface we conducted two online experiments. We found that for cases where data values are disproportionate as measured by normalized entropy and H-spread, wrapped bar charts allow participants to achieve higher accuracy on tasks of identification and ratio estimation but sometimes at the expense of more time spent and potentially more cognitive load. Finally, resulting from our focus group study, we develop a list of potential design improvements for such charts. Our findings, can serve as guidelines for visualizing datasets with disproportionate values using wrapped bar charts.


\bibliographystyle{SIGCHI-Reference-Format}
\bibliography{proceedings.bbl}

\end{document}